# Don't Just Teach, Explain! A Gamified 20Q Recommender for Cybersecurity Education

Mary Nusrat[1], Sarfuddin Bhuiyan[2], Gahangir Hossain[3]
[1, 3]Information Science, University of North Texas, Denton, TX, USA
[2]Mechanical Engineering, Bangladesh University of Eng. and Tech., BUET, Dhaka, Bangladesh
[1]NusratMary@my.unt.edu, [2]sbmehdi@gmail.com, [3]Gahangir.Hossain@unt.edu

*Abstract* — **The escalating complexity of modern cyber threats demands innovative approaches to security education that transcend traditional pedagogical methods. Conventional training paradigms often fail to engage learners meaningfully or develop the intuitive reasoning necessary for effective threat recognition. This paper introduces an interactive educational framework that reimagines cybersecurity awareness through the lens of a structured guessing game. Our approach integrates explainable artificial intelligence (XAI) principles with reinforcement learning to create a dynamic learning environment where users discover cybersecurity concepts through guided inquiry. The proposed system employs a policy-based reinforcement learning agent that assumes the role of a knowledgeable questioner, systematically narrowing down user-described security scenarios until it can both identify the underlying threat and provide transparent reasoning for its conclusion. By framing security education as an interactive dialogue, we transform passive knowledge acquisition into active discovery. We present the complete system architecture, detail the underlying algorithmic foundations, and demonstrate practical application through comprehensive case studies examining diverse attack vectors including the Cyber Kill Chain, phishing campaigns, ransomware outbreaks, and web application vulnerabilities. This work represents a significant departure from static security training methodologies, offering a personalized and game-based approach to cybersecurity education.**

## I. Introduction

The cybersecurity landscape is constantly evolving due to the introduction of new threats and vulnerabilities. The high volume of innovative security threats increases the difficulty of educating both the general public and professionals. Traditional trainings typically include lectures and reading static materials. These methods do not engage learners much and fall short of imparting the practical understanding and intuition required to counter highly sophisticated attacks.

To fill this gap, we propose an innovative solution that combines the captivating aspects of games with the powerful capabilities of explainable AI (XAI). The "20 Questions" (Q20) game, as a framework for this purpose, is a well-known exercise of logic and reasoning [5, 15]. Within our structure, a user briefly describes a cybersecurity concept as a brief text input to the game interface, and an AI agent in the role of a guesser tries to uncover it by asking detailed questions designed to narrow down the possibilities. This active process of learning not only retains attention but also facilitates the understanding of complex cybersecurity concepts.

Our work builds upon recent advancements in reinforcement learning for game-playing [1, 3] and explainable AI in cyber-awareness recommendation systems [2, 5]. We adapt the policy-based RL framework, which has been successfully applied to the Q20 game in guessing famous people [1] or product recommendation [2], to the domain of cybersecurity. Furthermore, this methodology is inspired by the work of Wu (2019) where a "Learning-to-Explain" (LTE) framework was successfully applied to product recommendations [2]. We adapt and extend this concept to the domain of cybersecurity, with an explanation mechanism from Wu (2019) [2], for a cybersecurity education recommender within the framework.

We outline the complete architecture of the system developed, which encompasses policy-based reinforcement learning and an entropy-based question selection/elimination policy, with an explanation generation module. To illustrate the framework's usefulness, we apply it to intricate cybersecurity topics, including the Cyber Kill Chain and multiple attack vectors, along with case studies. This approach is aimed at being a significant advancement in improving the effectiveness and reach of cybersecurity education.

The remainder of this paper is organized as follows. Section II reviews foundational literature related to LTE methodologies in cybersecurity and Q20 games. Section III introduces the proposed LTE framework with policy-based RL. Followed by Section IV, illustrating representative examples of attack vectors with case studies. Finally, Section VI concludes the study by summarizing key insights and implications.

## II. BACKGROUND

The application of game-based learning and artificial intelligence in education has gained significant traction in recent years. This section reviews existing literature in the areas of game-based learning for cybersecurity, the use of chatbots in education, and the principles of Explainable AI (XAI) that underpin our approach.

### *A. Game-Based Learning in Cybersecurity*

Game-based learning is widely recognized as an effective instructional strategy for enhancing student motivation and engagement, particularly when teaching complex or challenging concepts. In cybersecurity education, multiple studies have examined the use of games to simulate real-world scenarios and provide practical experience. These methods include tabletop card games and advanced digital simulations of cyber-attacks. For example, "Control-Alt-Hack" [16] is a card game developed to raise awareness about white-hat hacking, while online platforms, such as "Anti-Phishing Phil" [17], instruct users on defending against phishing attacks through interactive gameplay. These cases demonstrate the potential of gamification to make cybersecurity education more accessible and engaging. The proposed Q20 framework extends this approach by introducing a narrative-driven, deductive game that emphasizes conceptual understanding over technical simulation.

### *B. Chatbots for Educational Purposes*

Chatbots and conversational agents are increasingly being adopted as educational tools. These offer the possibility of individualized learning, immediate feedback on exercises, and natural responses to queries raised by learners. Research on security training using chatbots has yielded encouraging results. Although quantitative improvements in knowledge may not always reach statistical significance [3], the demonstrated willingness to learn and the increased engagement reported by users interacting with educational chatbots are invaluable. This approach utilizes a chatbot not just passively as an information repository but actively guiding the learners through structured dialog discovery and understanding complex cybersecurity issues [9, 12].

### *C. Explainable AI (XAI) in the Learning-to-Explain (LTE) Framework*

Explainable AI (XAI) is particularly important in this work and drives action in generating explanations. XAI aims to clarify and make comprehensible decisions and predictions of AI systems. This is particularly true in the case of cybersecurity, where explaining the reasoning behind an AI-powered decision could be crucial.

Wu (2019) [2] adapted the Learning-to-Explain (LTE) model, which is followed in our work. Wu demonstrated how a Q20-style game, combined with a question-ranking algorithm, is utilized for explainable product recommendations. The system recommends a product based on user responses and also explains the recommendation by revealing the most critical question-answer pair that impacted the decision. We apply this paradigm to cybersecurity education, where the objective is to guide the user in comprehending specific cyber-attack. In this framework, the "explanation" is a brief and precise description of the attack based on the user's path through the Q20 game. This methodology aligns with the broader goals of XAI, emphasizing transparency, interpretability, and user-centered understanding. In this case, the explanation is the outcome of engaging in an interactive educational process.

With the combination of game-based learning, educational chatbots, and explainable AI, the proposed framework offers a comprehensive solution to address the current gap in engaging cybersecurity education. It strives to develop an environment that would be just as transparent and empowering for the user while being effective and engaging. Figure 1 presents a root cause analysis that maps a strategic methodology aligned with enhancing cybersecurity education through the LTE-Q20 game framework.

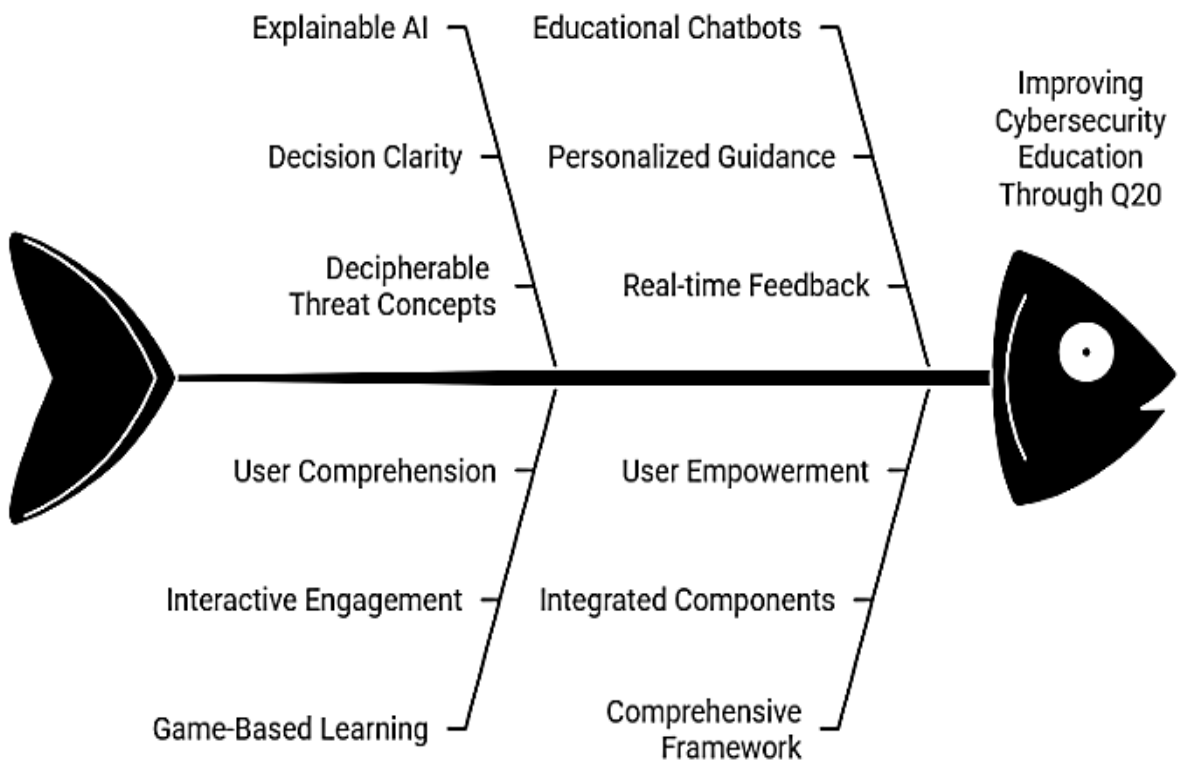

Fig 1. Ishikawa Diagram or Fishbone Analysis of Cybersecurity Awareness with Q20 Game.

## III. "LEARNING-TO-EXPLAIN" (LTE) FRAMEWORK

The proposed LTE framework for cybersecurity is designed as an interactive system where users engage with a question-based chatbot to learn about various cyber threats. This section details the architecture and the underlying algorithm of the framework.

The core of the framework is a "cybersecurity attack vectors", which include Phishing, Ransomware, and Man-in-the-Middle attacks. The "questions" are tailored to solicit details that would differentiate one attack from another. These questions focus on various endeavors of cyber-attacks, including the method of delivery, the goals of the attacker, the intended victim, and the possible consequences. The policy-based reinforcement learning is detailed in Figure 2. The interaction proceeds as follows:

**Initiation**: The user starts a session, which begins with a broad question like "Help me understand what a cyber-attack is".
**Questioning**: The bot scans its library of questions and ranks the question order with Shannon's entropy-based question elimination process, looking for the most relevant to the user's query in rank order. The user is then presented with a multiple-choice or yes/no question.
**User Response**: The user picks an answer from the selections provided to them.
**Probability Update**: The answer provided by the user elicits responses from the system in the form of probability estimates for all possible attack vectors. Attacks that rely on the user's answer are rewarded with a higher probability.
**Iteration**: The user is now presented with a new question, which is reprioritized in real time, and the entire cycle starts again.

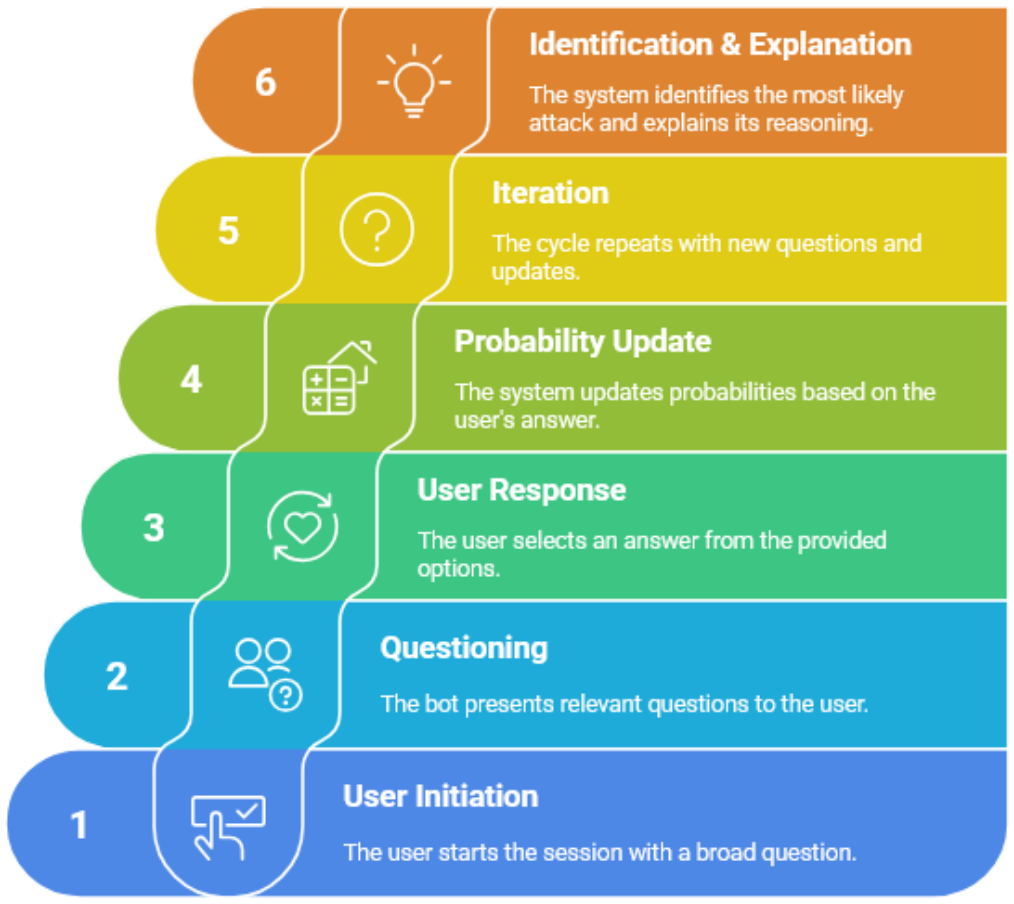

Fig 2. Training a Policy-based RL Agent in the LTE framework.

**Identification and Explanation:** The game ends when a set number of questions (up to 20) have been asked or when the likelihood of a single attack vector exceeds a specific threshold. The system then identifies the most probable attack and provides a detailed justification with an explanation. This explanation highlights how the user's decisions led to the identification of the particular threat and is closely linked to their responses.

The proposed framework is designed to be an interactive and adaptive learning environment. It consists of four primary components: the Knowledge Base, the Question Ranking approach, the Game Environment, and the Explanation Generation Module.

## *A. Knowledge Base*

A thorough knowledge base that encompasses a broad range of cybersecurity concepts forms the basis of our framework. The knowledge base is represented by Reference Matrix D, a matrix that links cybersecurity concepts and questions. Each entry in the matrix, denoted as $v_{m,n}^{i,l}$, represents the reference answer for concept $m$ and question $n$ for a given category $i$. We can think of it as $D$: A matrix where rows correspond to cybersecurity concepts and columns correspond to questions. Each cell $D[m][n]$contains the expected answer (or answers) for the concept $m$ to question $n$. In the case of multiple-choice questions, it will be a set of applicable options.

The matrix D is populated with data from established cybersecurity frameworks like the MITRE ATT&CK® framework [10] and the Cyber Kill Chain® [11], as well as from threat intelligence reports and academic research.

The entities in the knowledge base are associated with a set of attributes, which form the basis for the questions in the Q20 game. For instance, an entity like "Phishing" would have attributes, such as:
**Delivery Method:** Email, SMS, Social Media
**Objective:** Steal credentials, Deliver malware
**Primary Defense:** User awareness, Email filtering
**Associated Threat Actor Type:** Social engineer

## *B. Question Ranking Algorithms*

### *1. Policy-Based Reinforcement Learning Agent*

The core of our system is a policy-based RL agent that acts as the questioner in the Q20 game. The interaction between the user and the LTE framework is modeled as a finite-horizon Markov Decision Process (MDP). This setup enables the policy-based RL agent to sequentially choose questions (up to 20 questions) that optimize the long-term reward of accurately identifying the user's target cybersecurity concept [2]. The Q20 gaming process is defined by the MDP tuple:

$$\mathcal{M} = \langle \mathcal{S}, \mathcal{A}, P, R, \gamma \rangle$$

Here:
- $\mathcal{S}$: set of States (agent's belief over possible cyber-attack vectors)
- $t \in \{1,2,3, \ldots, 20\}$: Time step or turn index in the Q20 game
- $\mathcal{A}$: Action Space consists of the set of questions $Q = \{q_1, q_2, \ldots, q_N\}$. At each turn $t$, the agent selects a question $q_t \in Q$, according to the policy $\pi_\theta(a_t \mid s_t)$.
- $P(s'|s,a)$: Transition probability after receiving user response
- $R(s_t, a_t)$: Reward function (correct identification or intermediate information gain)
- $\gamma \in [0,1]$: The discount factor controls future reward weighting.

**State ($\mathbf{s} \in \mathcal{S}$):** The state is represented by a probability distribution over all the entities in the knowledge base, indicating the agent's current belief about the target concept [1]. The initial state can be a uniform distribution or a distribution [1] based on the prevalence of certain threats. Each state $s_t$ represents the probability distribution over cybersecurity concepts (attack vectors) given user responses up to time $t$ ($where\ t \leq 20$):

$$s_t = P(c \mid h_t) \quad (1)$$

Here,

- $s_t$: the state of the RL agent at time step $t$
- $c_m \in C$: Target cybersecurity concept or class (pre-stored in the knowledge base), e.g., Phishing, Ransomware, SQL Injection, Cross-Site Scripting, etc.
- $m = 1 \ldots M$ indicates the total number of cybersecurity concepts from the first one (m=1) to the last one (m=M)
- $h_t = \{(q_1, a_1), \ldots, (q_t, a_t)\}$: Dialogue history up to the time $t$
- $P(c \mid h_t)$: the probability distribution over possible concepts $c$, given that dialogue history

So, at each step $t$:

- The RL agent maintains a belief distribution $P(c \mid h_t)$ across all possible cyberattack vectors or concepts.
- Based on user responses (yes/no, multiple-choice, etc.), this belief is updated to reflect which cyberattack is most likely.

**Action (a):** An action is the selection of a question to ask the user. The set of all possible actions is derived from the attributes in the knowledge base [2]. The RL agent chooses a question $q_t \in \mathcal{A}$, according to a stochastic policy $\pi_\theta(a_t \mid s_t)$:

$$Action\ selection: a_t \sim \pi_\theta(a_t | s_t) \quad (2)$$

here $\pi_\theta$ is parameterized by a neural network that maps the belief state $s_t$ to a distribution over possible questions.

**Policy (π(a|s)):** The agent's policy is a neural network that takes the current state as input and outputs a probability distribution over the available questions. This stochastic policy allows for a diverse range of questions, making the game more engaging and robust to noisy or uncertain user answers [1]. The agent maximizes expected cumulative reward:

$$J(\theta) = \mathbb{E}_{\pi_\theta} \sum_{t=1}^{T} \gamma^{t-1} r_t \quad (3)$$

here $r_t$ is the immediate reward at step $t$ (estimated via RewardNet or information gain). Also, $\mathbb{E}_{\pi_\theta}$ is the expectation (average) taken over all possible action sequences generated by the current policy $\pi_\theta$.

Following the policy gradient theorem (Williams, 1992; Wu, 2019), the gradient of $J(\theta)$ is:

$$\nabla_\theta J(\theta) = \mathbb{E}_{\pi_\theta}[\sum_{t=1}^{T} \nabla_\theta \log \pi_\theta(a_t \mid s_t)\, G_t] \quad (4)$$

here $G_t = \sum_{k=t}^{T} \gamma^{k-t} r_k$ is the discounted return from time step $t$.

In implementation, a baseline $b_t$ (e.g., moving average of rewards) is often subtracted to reduce variance:

$$\nabla_\theta J(\theta) = \mathbb{E}_{\pi_\theta}[\sum_{t=1}^{T} \nabla_\theta \log \pi_\theta(a_t \mid s_t)\, (G_t - b_t)] \quad (5)$$

**Reward (r):** A direct reward is given only at the end of the game: a positive reward for a correct guess and a negative reward for an incorrect guess [2, 12]. To address the issue of sparse rewards, we employ a "RewardNet" that estimates an immediate reward for each question asked, based on its ability to narrow down the potential targets [2, 13]. Wu's LTE model used a sparse terminal reward for correct product identification. In this adaptation, the reward combines end-game accuracy and information gain per question:

$$r_t = \alpha\, \Delta H_t + \beta\, R_{\text{final}} \quad (6)$$

here:

- $\Delta H_t = H(s_{t-1}) - H(s_t)$ is the reduction in Shannon entropy (information gain)
- $R_{\text{final}} = \begin{cases} +1, & \text{if correct attack vector identified} \\ -1, & \text{otherwise} \end{cases}$
- $\alpha, \beta$ are balancing coefficients.
- The update factor $\beta$ depends on comparing user's actual answer with the reference answers in matrix D

**Training:** The agent is trained through continuous interaction with a user simulator, which is programmed to think of a cybersecurity concept and answer questions based on the knowledge base. The policy network is updated using the REINFORCE algorithm to maximize the expected long-term reward [2].

The reinforcement learning model is trained through self-play, where the system simulates thousands of Q20 games against itself. This enables the policy network to learn the complex relationships between questions, answers, and attack vectors, and to develop sophisticated, long-term strategies for guiding users to the correct identification [5].

The initial phase of the Q20 game can be improved using an entropy-based method. The goal is to select questions that provide the most information gain, regardless of the user's answer. This is achieved by calculating the "negative Shannon entropy" for each possible question. Questions that most effectively decrease uncertainty across potential attack vectors are ranked higher. This approach ensures that the initial questions are broad and quickly narrow down the list of possibilities.

### 2. *Entropy-Based Question Selection*

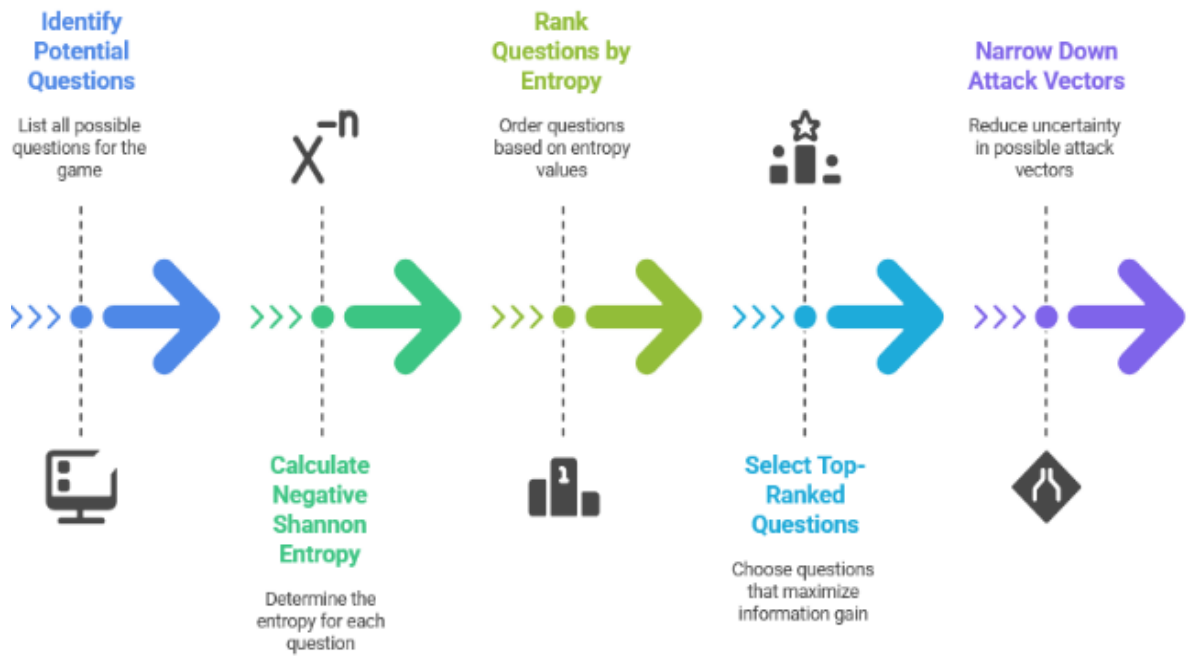

Fig 3. Entropy-based question ranking process.

Initial Belief State (Prior Distribution): Equation (7) initializes the system's belief about which cybersecurity concept the user is thinking of:

$$w'(c_m) = \frac{w(c_m)}{\sum_{m=1}^{M} w(c_m)} \tag{7}$$

here,

- $m = 1, \dots M$: is the total number of cybersecurity concepts (attack vectors) in the knowledge base
- $w(c_m)$: Prior weight of concept $c_m$
- $w'(c_m)$: Normalized probability

Answer Likelihood: Equation (8) calculates how a user's answer to a question aligns with a given cybersecurity concept.

$$Y_{m,n}^{l} = \frac{f_{m,n}^{l} + \alpha \mathrm{I}(v_{m,n}^{l} = \text{"yes"})}{\sum_{l=1}^{L_n} \{f_{m,n}^{l} + \alpha \mathrm{I}(v_{m,n}^{l} = \text{"yes"})\}} \tag{8}$$

- $q_n$: A specific question
- $l$: A specific answer option to question $q_n$
- $f_{m,n}^{l}$: Historical frequency of users selecting option $l$ for question $q_n$ when the target was $c_m$
- $v_{m,n}^{l}$: Reference answer from the knowledge base
- $\alpha$: A balancing parameter between historical user data ($\alpha = 0$) and the reference knowledge base ($\alpha$ is large)
- $\mathrm{I}(\cdot)$: Indicator function
- $Y_{m,n}^{l}$: The likelihood that a user thinking of concept $c_m$ would answer with option $l$ to question $q_n$

Entropy-Based Question Ranking: This is used for the entropy-based question selection method, ranking questions by their expected information gain.

$$w'(q_n) = \sum_{m=1}^{M} w'(c_m) \times (-H(s)) \tag{9}$$

here,

- $w'(q_n)$: The weight of question $q_n$. A higher (less negative) value means the question is more informative.
- $H(S) = \sum_{l=1}^{L_n} Y_{m,n}^{l} \log_2 Y_{m,n}^{l}$ is Shannon's entropy. A question that minimizes the expected entropy (i.e., reduces uncertainty the most) is preferred.

Information Gain for Specific Concept (Probability Jump): Equation (10) quantifies how much a user's answer shifts the system's belief, which is crucial for generating the final explanation. From the Entropy-Based Method:

$$g_{t+1}(c_m, q_t, a) = w'_{t+1}(c_m) - w'_t(c_m) \tag{10}$$

- $g(\cdot)$: Information gain or "probability jump"
- $q_t, a$: The question asked at turn $t$ and the user's answer
- $w'_t(c_m), w'_{t+1}(c_m)$: The probability of concept $c_m$ before and after the update.

Belief State Update in Reinforcement Learning: In the policy-based RL approach, the entire belief state vector $s_t$ is updated. The information gain is computed as the difference between successive states.

$$g_{t+1}(\{c_m\}, q_t, a) = s_{t+1} - s_t \tag{11}$$

- $s_t, s_{t+1}$: Belief state vectors before and after user's answer

After the user answers, the system updates its beliefs and tracks how much each answer contributed to identifying the final concept (10 and 11). This "information gain" is used to generate the final, human-readable explanation for the recommendation.

## C. *Game Environment and User Interaction*

The game environment facilitates the interaction between the user and the RL agent. The user initiates the game by articulating a general description of a cybersecurity concept in a few sentences. The agent then asks a series of yes/no or multiple-choice questions to the user. The user's answers update the agent's belief state, which in turn informs the selection of the next question. The game will be deployed as a chatbot, a web-based application, or can be integrated into existing learning management systems.

### 1. *State Representation and Update*

The dialog state $s_t$ is a vector that encapsulates the current belief over all cybersecurity concepts, initialized with a prior distribution (e.g., uniform or based on threat prevalence). After the agent asks a question $q_t$ and receives the user's answer $a_t$ (a set of selected options $I \subset [1, \dots, L_{q_t}]$), the state is updated as follows:

$$s_{t+1} = \text{normalize}(s_t \odot \beta)$$

Here, $\odot$ denotes the element-wise product, and $\beta = [Y_{1,q_t}^{(I)}, \dots, Y_{M,q_t}^{(I)}]$ is a vector where each element $Y_{m,q_t}^{(I)} = \sum_{l \in I} Y_{m,q_t}^{l}$ aggregates the likelihoods from Equation (8) for the user's selected options. Normalization ensures the state vector remains a valid probability distribution.

### 2. *LTE-RewardNet and Training Objective*

To address the challenge of sparse rewards and to incorporate historical interaction data, we employ a neural network-based LTE-RewardNet, denoted by $R$ [2]. This network estimates the immediate reward $r_{t+1}$ by taking a tuple $\langle s_t, q_t, a_t, c_{\text{target}}\rangle$ as input, where $c_{\text{target}}$ is the ground-truth concept.

The agent's policy $\pi_\theta(a_t \mid s_t)$ is a neural network parameterized by $\theta$ that maps the current state to a probability distribution over questions. The objective is to maximize the expected cumulative discounted reward from equation (3). The policy is optimized using the REINFORCE algorithm with a baseline (provided by a value network $V$) to reduce variance [2].

### 3. *Training Algorithm*

The joint training process for the LTE-RewardNet ($R$), the value network ($V$), and the policy network ($\pi_\theta$) is detailed in Algorithm. It proceeds through self-play, where the agent interacts with a user simulator that answers questions based on a pre-defined knowledge base. In the RL process, an episode is defined as a quadruple $\langle s, qt, r_{t+1}, s'\rangle$. In the time step t, the long-term reward $R_t$ is traditionally defined to be $R_t = \sum_{k=1}^{T-t-1} \gamma^{k} r_{k+t+1}$. The Algorithm shows the steps of training details for the LTE framework.

**Algorithm:** Joint Training of LTE-RewardNet, Policy, and Value Networks for Cybersecurity Q20

**Input:**

- Cybersecurity concepts $C = \{c_1, \dots, c_M\}$
- Questions $Q = \{q_1, \dots, q_N\}$
- Reference Matrix $D$(where $D[m][n]$gives the reference answer for concept $c_m$ and question $q_n$)
- User Simulator (to provide answers based on a target concept)

**Output:** Trained Policy Network $\pi_\theta$, Reward Network $R$, Value Network $V$

1. Initialize episode memory $\mathcal{M} \leftarrow \emptyset$
2. Randomly initialize parameters for $R$, $V$, and $\pi_\theta$, for the user's brief text input
3. **for** epoch = 1 **to** $Z$**do**
4. Sample a target concept $c_{target}$from $C$(weighted by prior $\mathbf{w}(c_m)$)
5. Initialize dialog state $s_0$(e.g., uniform or prior distribution over $C$)
6. Initialize sets $S_1 \leftarrow \emptyset$, $S_2 \leftarrow \emptyset$
7. **for** turn $t$= 1 **to** $T$(max turns, e.g., 20) **do**
8. Select a question $q_t \sim \pi_\theta(\cdot \mid s_{t-1})$
9. Obtain user's answer $a_t$from User Simulator (based on $c_{target}$and $D$)
10. Update belief state $s_t = s_{t-1} \odot \beta$// $\beta$depends on $a_t$and $D$(using Eq. 2)
11. Store $(s_t, q_t, a_t)$in $S_1$
12. **if** argmax$(s_t) = c_{target}$**then break**
13. **end for**
14. **if** episode ended with correct identification **then** $r_T = +\kappa$**else** $r_T = -\kappa$
15. **for** each $(s_t, q_t, a_t)$in $S_1$**do**
16. Compute immediate reward $r_{t+1} \leftarrow R(\langle s_t, q_t, a_t, c_{target}\rangle)$
17. Store $(s_t, q_t, a_t, r_{t+1})$in $S_2$
18. **end for**
19. **for** each $(s_t, q_t, a_t, r_{t+1})$in $S_2$**do**
20. Compute discounted cumulative reward: $r'_{t+1} = \text{sigmoid}(\sum_{k=0}^{T-t-1} \gamma^k r_{t+k+1})$
21. Update $V$ by $(s_t, r'_{t+1})$// Train Value Network to predict $r'_{t+1}$
22. $v_{t+1} \leftarrow V(s_t)$
23. Store $(s_t, q_t, a_t, c_{target}, r'_{t+1} - v_{t+1})$in $\mathcal{M}$
24. **if** | $\mathcal{M}$ |> BatchSize **then**
25. Update $R$ and $\pi_\theta$using mini-batches from $\mathcal{M}$
26. **end if**
27. **end for**
28. **end for**

### D. *Information Gain and Explanation Generation Module*

A key feature of the LTE framework is its ability to provide clear and contextualized explanations. This is achieved by tracking the "information gain" associated with each question-answer pair. The information gain is defined as the change in probability of the final identified attack vector after the user answers a particular question [2].

After the game concludes, the system examines all the dialogues and determines which question-answer combination provided the most information. The final explanation is then built around this "pivotal" piece of information (for instance, "Based on your answer that the attack arrived via a deceptive email, the most likely threat is Phishing"), will be stated explicitly, if the user's response to a question regarding the attack's delivery mechanism significantly increases the likelihood of "Phishing". The learning process is transparent and individualized due to the direct connection between the user's input and the derived system output [6, 7].

## IV. Attack Vectors and Case Studies

This framework can be used to explain a varied array of complex cybersecurity topics. Below are a few examples:

### A. *The Cyber Kill Chain*

The Cyber Kill Chain is a model developed by Lockheed Martin that outlines the stages of a cyber-attack [11]. Explaining this multi-stage process can be challenging. Our Q20 game can deconstruct it by focusing on the distinct phases.

**Sample Q20 Dialogue for the "Reconnaissance" Stage:**

**Agent:** Briefly describe the incident you are experiencing
**User:** … (Brief text input)
**Agent:** Is the activity you're thinking of, is the very first stage of a planned cyber-attack?
**User:** Yes.

| |
|---|
| **Agent:** Does this stage involve actively probing the target's network for vulnerabilities?<br>**User:** No.<br>**Agent:** Is the primary goal of this stage to gather information about the target, such as employee names or network infrastructure details?<br>**User:** Yes.<br>**Agent:** Are techniques like harvesting email addresses or researching social media profiles common in this stage?<br>**User:** Yes.<br>**Agent:** I believe the stage you are thinking of is **Reconnaissance**. |
| **Generated Explanation:** "Reconnaissance is the initial phase of a cyber-attack where the attacker gathers information about the target. This is often a passive stage that doesn't involve direct engagement with the target's systems. The goal is to identify potential vulnerabilities and plan the subsequent stages of the attack." |

*B. Attack Vectors Explained with Case Studies*

Attack vectors refer to the techniques employed by threat actors to obtain unauthorized access to networks or systems. The proposed framework facilitates user comprehension of these vectors through interactive questioning supported by real-world examples. The inclusion of case studies, followed by the application of the framework, illustrates how the approach can effectively contextualize and reinforce key cybersecurity concepts.

### 1. PHISHING

**Case Study: The 2016 email leak from the DNC**

In 2016, spear-phishing email attacks were made to the members of the Democratic National Committee (DNC) claimed as Google security alerts, asking them to change their passwords. A few employees unknowingly allowed the hackers to access their email accounts by complying. Eventually, thousands of private emails were stolen and made public with very dire political consequences [18].

**Sample Q20 Dialogue for Phishing cyberattack:**

| |
|---|
| **Agent:** Briefly describe the incident you are experiencing<br>**User:** I received an email asking me to verify my account. (Text input)<br>**Agent:** How is the attack delivered?<br>(A: Deceptive email, B: Malicious software, C: Network interception)<br>**User:** A.<br>**Agent:** What is the primary goal of the attack? (A: Steal login credentials, B: Encrypt files for ransom, C: Disrupt service)<br>**User:** A.<br>**Agent:** Did the email appear from a trusted organization (e.g., your bank, university, or IT department)?<br>**User:** Yes.<br>**Agent:** When you hovered over the link, did the URL look suspicious or different from the official website?<br>**User:** Yes.<br>**Agent:** Did the message contain a link or attachment asking you to log in or download something?<br>**User:** Yes.<br>**Agent:** Did you or anyone else enter credentials or download a file from the email?<br>**User:** No.<br>**Agent:** Have you noticed any unusual account activity or unauthorized access since then?<br>**User:** No.<br>**Agent:** Did this attack result in the theft and publication of sensitive emails?<br>**User:** Yes.<br>**Agent:** The attack vector you are thinking of is **Phishing**. |
| **Generated Explanation:** Phishing is a social engineering attack in which adversaries convince users to reveal sensitive data. Most often, this happens through email or text message. In the 2016 DNC hack, the assailants used spear-phishing emails purportedly from Google to obtain user credentials. Once those logins were obtained, access was granted to DNC email accounts; later, thousands of extremely sensitive communications were published. This example shows how phishing bypasses human trust more than any technical vulnerability but can have devastating effects both organizationally and politically. |

### 2. RANSOMWARE

**Case Study: The WannaCry Attack (2017)**

The WannaCry ransomware attack happened in 2017 and quickly propagated to over 150 countries, affecting hundreds of thousands of computers. It took advantage of a flaw within old Windows systems (EternalBlue) to inject ransomware that would then lock users' files. A ransom note would pop up asking for payments to be made in Bitcoin so that keys could be sent to unlock the files. This attack brought about a huge disruption all over the world with business hours lost; it particularly affected the UK's National Health Service (NHS) and was one of the most severely impacted, where appointments were canceled and treatments delayed, besides considerable financial loss [19].

**Sample Q20 Dialogue for Ransomware cyber-attack:**

| |
|---|
| **Agent:** Briefly describe the incident you are experiencing<br>**User:** My computer suddenly showed a message saying my files are encrypted. (Text input)<br>**Agent:** What is the primary impact of the attack on the victim's data?<br>(A: Data is stolen, B: Data is encrypted and made inaccessible, C: Data is publicly exposed)<br>**User:** B. |

**Agent:** What is the attacker's primary motivation? (A: Financial gain through a ransom demand, B: Espionage, C: Political activism)
**User:** A.
**Agent:** Data encryption and demanding payment - are these part of the attack?
(A: Yes, a ransom note and encryption are involved; B: No, the attacker merely steals data; C: The attacker deletes data without requesting a ransom.)
**User: A.**
**Agent:** What must victims do to regain access to their encrypted files?
(A: Pay a ransom in Bitcoin, B: Reset their passwords, C: Contact the attacker directly)
**User:** A.
**Agent:** The attack vector you are thinking of is **Ransomware**.

**Generated Explanation:** Ransomware is a type of malware that encrypts a victim's data and demands payment, often in cryptocurrency, to restore access. The 2017 WannaCry attack is one of the most well-known examples. It leveraged the EternalBlue vulnerability in unpatched Windows systems to spread rapidly across networks. Victims were locked out of their files and pressured to pay a ransom in Bitcoin for decryption keys. The attack had devastating real-world effects, particularly on the UK's National Health Service (NHS), where critical medical operations were disrupted. WannaCry illustrates how ransomware combines technical exploitation with financial extortion, making it one of the most damaging and costly forms of cyberattacks to date.

### 3. SQL Injection (SQLi)

**Case Study: The 2014-2015 TalkTalk Hack**

In 2015, the UK telecommunications company TalkTalk suffered a major data breach that exposed the personal details of over 150,000 customers. The primary attack vector was an SQL injection vulnerability on their website [20].

**Sample Q20 Dialogue for SQL Injection:**

**Agent:** Briefly describe the incident you are experiencing
**User:** Our website started showing strange database errors, and some user information got exposed after someone entered unusual characters into the login form. (Text input)
**Agent:** Does this attack target a web application?
**User:** Yes.
**Agent:** Does it involve manipulating the input fields of a website, such as a search bar or login form?
**User:** Yes.
**Agent:** Is the goal to trick the application into executing unintended database commands?
**User:** Yes.
**Agent:** Can this vulnerability expose sensitive information such as usernames, passwords, or financial data from the database?
**User:** Yes.
**Agent:** Is this attack possible due to improper validation or sanitization of user inputs in the application's code?
**User:** Yes.
**Agent:** Does the attacker typically insert malicious SQL statements into input fields to gain unauthorized access to data?
**User:** Yes.
**Agent:** The attack vector you are thinking of is **SQL Injection**.

**Generated Explanation:** "SQL Injection (SQLi) is a web security vulnerability that allows an attacker to interfere with the queries that an application makes to its database. It generally allows an attacker to view data that they are not normally able to retrieve. The TalkTalk breach is a prime example of how damaging this vulnerability can be".

### 4. Cross-Site Scripting (XSS)

**Case Study: The 2018 British Airways Data Breach**

In 2018, British Airways' website and mobile app were compromised, leading to the theft of financial data from approximately 380,000 customers. The attack was carried out by injecting malicious code into the website, a classic example of a cross-site scripting attack [21].

**Sample Q20 Dialogue for Cross-Site Scripting:**

**Agent:** Briefly describe the incident you are experiencing
**User:** When a user visits our website, malicious pop-ups appear and some pages automatically redirect to other websites. (Text input)
**Agent:** Does this attack involve injecting malicious scripts into a trusted website?
**User:** Yes.
**Agent:** Is the malicious script executed in the victim's web browser?
**User:** Yes.
**Agent:** Can this attack be used to steal session cookies or hijack user accounts?
**User:** Yes.
**Agent:** Does this attack rely on manipulating server-side database queries?
**User:** No.
**Agent:** Does it typically involve direct access to backend systems or file servers?
**User:** No.
**Agent:** Is this attack often carried out through vulnerable input fields, comment sections, or URL parameters that fail to properly sanitize user input?
**User:** Yes.
**Agent:** The attack you are thinking of is **Cross-Site Scripting (XSS)**.

**Generated Explanation:** "Cross-Site Scripting (XSS) is a type of security vulnerability typically found in web applications. XSS attacks enable attackers to inject client-side scripts into web pages viewed by other users. The British Airways breach demonstrated how XSS can be used to skim payment card information in real-time."

## V. Discussion and Future Work

The "Learning to Explain Cybersecurity through a Q20 Game" framework offers several advantages over traditional cybersecurity education methods:

**Engagement:** The gamified approach fosters active participation and makes learning more enjoyable.

**Adaptability:** The RL agent can tailor the line of questioning to the user's level of understanding.

**Explainability:** The framework provides clear and contextualized explanations based on interactive dialogue.

**Scalability:** The knowledge base can be continuously expanded to include contemporary and evolving vulnerabilities.

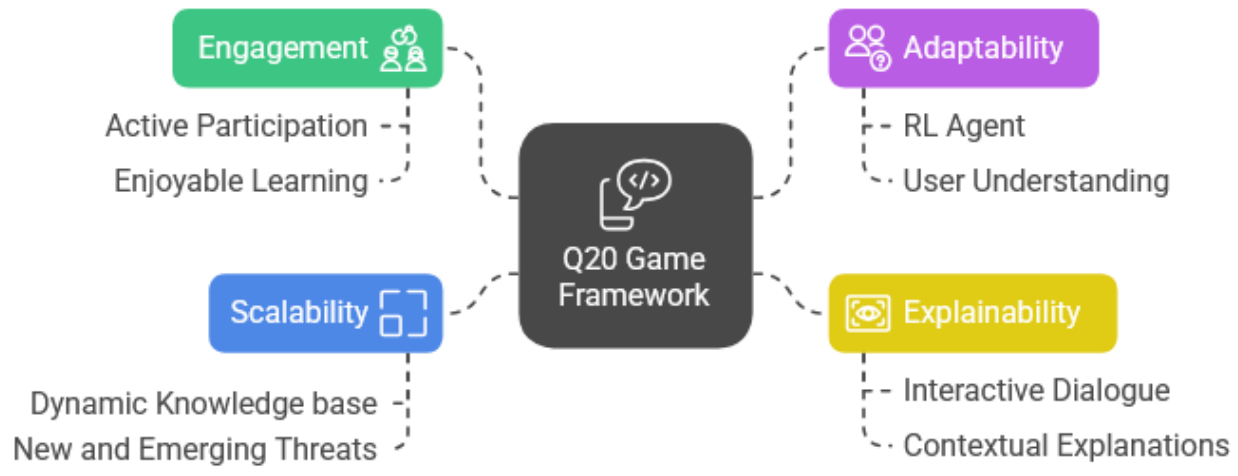

Fig 4. Advantages of the LTE Cybersecurity Framework using Q20 Gaming.

At every stage of the learning process, the Entropy-based Question Selection module uses question ranking to select the best questions [24]. This part determines the expected information gain for each possible question based on the concepts of information theory, making sure that learners get the most out of their education while lowering the number of questions needed to meet learning goals [14].

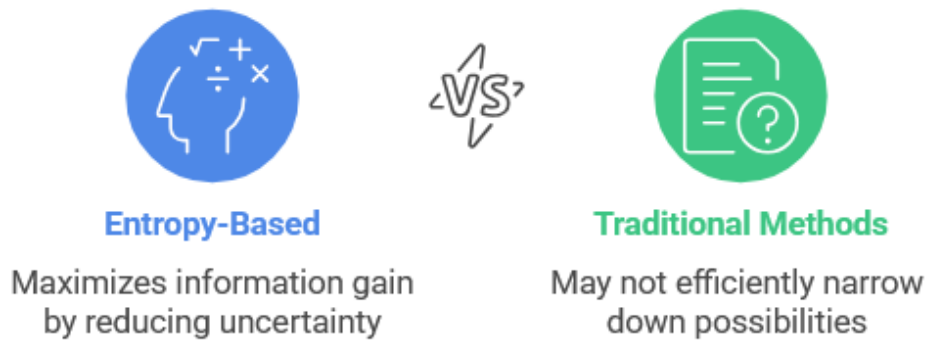

Fig 5. Shannon's entropy-based Question ranking vs. the traditional way.

However, the framework has its limitations. The quality of the learning experience largely depends on the detail and accuracy of the knowledge base that provides contextual explanations to learners. In an ideal system, the knowledge base would pre-store all types of cyber threats with extensive data storage. Additionally, developing a strong and well-trained RL agent requires significant computational resources and training data.

While the methodology outlined by Wu [2] serves as the foundation for this study, the proposed LTE framework introduces an attention-based explainability mechanism for implementation. The most logical and synergistic module to integrate with the existing RL policy in the Q20 game is an Attention-based mechanism, specifically leveraging the Information Gain/Probability Jump concept inherent in the original framework. While SHapley Additive exPlanations (SHAP) and Local Interpretable Model-agnostic Explanations (LIME) are powerful post-hoc feature attribution methods - both primarily providing local interpretability, with SHAP further enabling aggregated global explanation, an Attention-based approach offers the advantage of being inherently integrated within the RL agent's decision-making process. This integration ensures that the generated explanations are both contextually meaningful and pedagogically relevant to the game's interactive flow.

The proposed Attention-based explainability module will be integrated into the policy-based RL framework to improve transparency and interpretability in the agent's decision-making. By dynamically highlighting key question-answer interactions, the attention mechanism will support dialogue-level interpretability, allowing for a clearer understanding of the reasoning process behind the agent's actions.

## VI. Conclusion

This article introduces a novel framework that combines 20 Questions (Q20) and Learning-to-Explain (LTE) with cybersecurity, creating an interactive and explainable cybersecurity training tool. By using the Q20 model, this approach transforms the learning experience into an interactive conversation where users, guided by a question-based game environment, identify and understand various cyber threats. This method enhances user engagement and provides meaningful, context-aware explanations of complex cybersecurity concepts. The paper details the design of the framework, demonstrates its practical operation in justifying complex attack vectors (through case studies), and explains the underlying principles of RL that optimize the question-asking process. The system design offers a scalable, practice-oriented approach to cybersecurity education.

By carrying the Q20 gaming format, this method makes education an engaging conversation. Users can identify and understand different cyberthreats within a game environment. Along with improving user engagement with cognitive retention, this approach provides clear, relevant explanations of complex cybersecurity ideas.

Initially, the process begins with the user briefly articulating a concept and then engaging the agent by asking what it is thinking of. The average number of questions required for convergence during self-play simulations, before the Q20 agent successfully identifies the cyberthreat type and presents the generated explanation to the player, is determined by the policy-based Reinforcement Learning (RL) agent, which functions as a diagnostic cyber-threat identifying mechanism.

For future research, the game-based learning framework will undergo a pilot evaluation to substantiate its effectiveness with empirical data. As part of the extended work, accuracy will be introduced as an additional metric to assess user acceptance and learning outcomes within this explainable, game-based cybersecurity learning environment.

## Acknowledgement

This research was conducted with the support of the Office of Naval Research (ONR), U.S.A., under Award No. N00014-23-1-2454.